
\input harvmac

\overfullrule=0pt




\def\np {  Nucl. Phys. }
\def \pl { Phys. Lett. }
\def \mpl { Mod. Phys. Lett. }
\def \prl { Phys. Rev. Lett. }

\def \cmp { Commun. Math. Phys. }

\def\b{\beta}

\def\g{\gamma}

\def\d{\delta}
\def\D{\Delta}
\def\e{\epsilon}

\def\no{\noindent}
\def\hb{\hfill\break}
\def\qq{\qquad}
\def\bl{\bigl}
\def\br{\bigr}

\def\IR{\relax{\rm I\kern-.18em R}}
\def\Z{\bf Z}



 \lref\BSeffaction{I. Bars and K. Sfetsos,  Phys. Rev. {\bf D48} (1993) 844.}
\lref\tseytlin{A.A. Tseytlin, preprint CERN-TH.6804/93.}
 \lref\IBhet{ I. Bars, Nucl. Phys. {\bf B334} (1990) 125. }
 \lref\BN{ I. Bars and D. Nemeschansky, Nucl. Phys. {\bf B348} (1991) 89.}
 \lref\IBCS{ I. Bars,``Curved Spacetime Strings and Black Holes", in Proc.
   {\it XX$^{th}$ Int. Conf. on Diff. Geometrical Methods in Physics}, eds. S.
   Catto and A. Rocha, Vol. 2, p. 695, (World Scientific, 1992).}
 \lref\WIT{E. Witten, Phys. Rev. {\bf D44} (1991) 314.}
 \lref\CRE{M. Crescimanno, Mod. Phys. Lett. {\bf A7} (1992) 489.}
 \lref\HOHO{J. B. Horne and G. T. Horowitz, Nucl. Phys. {\bf B368} (1992) 444.}
 \lref\BSthree{I. Bars and K. Sfetsos, Mod. Phys. Lett. {\bf A7} (1992) 1091.}
 \lref\BShet{I. Bars and K. Sfetsos, Phys. Lett. {\bf 277B} (1992) 269.}
 \lref\FRA{E. S. Fradkin and V. Ya. Linetsky, Phys. Lett. {\bf 277B} (1992)
73.}
 \lref\ISH{N. Ishibashi, M. Li, and A. R. Steif,
         Phys. Rev. Lett. {\bf 67} (1991) 3336.}
 \lref\HOR{P. Horava, Phys. Lett. {\bf 278B} (1992) 101.}
 \lref\RAI{E. Raiten, ``Perturbations of a Stringy Black Hole'',
         Fermilab-Pub 91-338-T.}
 \lref\GER{D. Gershon, ``Exact Solutions of Four-Dimensional Black Holes in
         String Theory'', TAUP-1937-91.}
 \lref\others{ M. Crescimanno, Mod. Phys. Lett. {\bf A7} (1992) 489;\semi
   {J. B. Horne and G. T. Horowitz, Nucl. Phys. {\bf B368} (1992) 444};\semi
   {E. S. Fradkin and V. Ya. Linetsky, Phys. Lett. {\bf 277B} (1992) 73};\semi
   {P. Horava, Phys. Lett. {\bf 278B} (1992) 101};\semi
   {E. Raiten, ``Perturbations of a Stringy Black Hole'', Fermilab-Pub 91-338-
     T};  \semi
   {D. Gershon, ``Exact Solutions of Four-Dimensional Black Holes in
         String Theory'', TAUP-1937-91.} }
 \lref \GIN{P. Ginsparg and F. Quevedo,  Nucl. Phys. {\bf B385} (1992) 527. }
 \lref\BSglo{I. Bars and K. Sfetsos, Phys. Rev. {\bf D46} (1992) 4495.}
 \lref\BSexa{I. Bars and K. Sfetsos, Phys. Rev. {\bf D46} (1992) 4510.}
 \lref\SFET{K. Sfetsos, ``Conformally Exact Results for
    $SL(2,\IR)\otimes SO(1,1)^{d-2} /SO(1,1)$ Coset Models'',
    USC-92/HEP-S1 (hep-th/9206048), to appear in Nucl. Phys. }
 \lref\groups{
    M. Crescimanno. Mod. Phys. Lett. {\bf A7} (1992) 489. \hb
    J. B. Horne and G.T. Horowitz, Nucl. Phys. {\bf B368} (1992) 444. \hb
    E. S. Fradkin and V. Ya. Linetsky, Phys. Lett. {\bf 277B} (1992) 73. \hb
    P. Horava, Phys. Lett. {\bf 278B} (1992) 101.\hb
    D. Gershon, TAUP-1937-91 }
 \lref\KASU{Y. Kazama and H. Suzuki, Nucl. Phys. {\bf B234} (1989) 232 \semi
    Phys. Lett. {\bf 216 B} (1989) 112.}
 \lref\NAWIT{C. Nappi and E. Witten, ``A Closed, Expanding Universe in String
           Theory'', IASSNS-HEP-92/38.}
 \lref\BSslsu{I. Bars and K. Sfetsos $SL(2,\IR)\times SU(2)/R^2$ String Model
   in Curved Spacetime and Exact Conformal Results",
   Phys. Lett. B301 (1993) 183. }
\lref\WITanom{E. Witten, Comm. Math. Phys. {\bf 144} (1992) 189.}
\lref\IBhetero{I. Bars, Phys. Lett {\bf 293B} (1992) 315.}
\lref\IBerice{I. Bars, ``Superstrings on Curved Spacetimes", USC-92/HEP-B5
   (hep-th 9210079), in {\it String Quantum Gravity and Physics at the
   Planck Energy Scale}, Ed. N. Sanchez, World Scientific (1993), page. }
\lref\DVV{R. Dijgraaf, E. Verlinde and H. Verlinde, Nucl. Phys. {\bf B371}
   (1992) 269}
 \lref\TSEY{A.A. Tseytlin, Phys. Lett. {\bf 268B} (1991) 175.
      \semi I. Jack, D. R. T. Jones and J. Panvel, ``Exact Bosonic and
         Supersymmetric String Black Hole Solutions", LTH-277.}
 \lref\BST { I. Bars,  K. Sfetsos and A.A. Tseytlin, unpublished. }
 \lref\TSEYT{ A.A. Tseytlin, ``Effective Action in Gauged WZW Models and Exact
   String Solutions", Imperial/TP/92-93/10.}
 \lref\SHIF { M.A. Shifman, Nucl. Phys. {\bf B352} (1991) 87.}
 \lref\SHIFM { H. Leutwyler and M.A. Shifman,
                    Int. J. Mod. Phys. {\bf A7} (1992) 795. }
 \lref\POLWIG { A.M. Polyakov and P.B. Wiegman, Phys. Lett. {\bf 131B} (1984)
                             121.  }
 \lref\BCR{K. Bardakci, M. Crescimanno and E. Rabinovoici,
               Nucl. Phys. {\bf B344} (1990) 344. }
 \lref\GWZW{ E. Witten, \np {\bf B223} (1983)422;
         \semi   K. Bardakci, E. Rabinovici andB. S\"aring,
                        Nucl. Phys. {\bf B299} (1988) 157;
         \semi K. Gawedzki and A. Kupiainen, Phys. Lett. {\bf 215B} (1988) 119;
             Nucl. Phys. {\bf B320} (1989) 625. }
 \lref\SCH{ D. Karabali, Q-Han Park, H.J. Schnitzer and Z.Yang,
                   Phys. Lett. {\bf B216} (1989) 307;
    \semi  D. Karabali and H.J. Schnitzer, Nucl. Phys. {\bf B329} (1990) 649. }

 \lref\KIR{E. Kiritsis, Mod. Phys. Lett. {\bf A6} (1991) 2871. }

 \lref \GKO{K. Bardakci and M.B. Halpern, Phys. Rev. D3 (1971) 2493;
         \semi  P. Goddard, A. Kent and D. Olive, Phys. Lett. 152B (1985) 88.}

 \lref\KAC{V.G. Kac, Func. Anal. App. {\bf 1} (1967) 238; \semi
          R.V. Moody, Bull. Am. Math. Soc. {\bf 73} (1967) 217. }

 \lref\WITT{ E. Witten, Comm. Math. Phys. {\bf 92} (1984) 455.}


\lref \sen {A. Sen, preprint TIFR-TH-92-57. }
\lref \zam  { Al. B. Zamolodchikov, preprint ITEP 87-89. }
\lref \tse { A.A. Tseytlin, \pl {\bf 264B} (1991) 311. }
\lref  \kumar  { M. Ro\v cek and E. Verlinde, \np B373(1992)630; A. Kumar,
   preprint CERN-TH.6530/92;
   S. Hussan and A. Sen,  preprint  TIFR-TH-92-61;  D. Gershon,
   preprint TAUP-2005-92; X. de la Ossa and F. Quevedo, preprint NEIP92-004; E.
   Kiritsis, preprint LPTENS-92-29. }
\lref \rocver { A. Giveon and M. Ro\v cek, \np B380(1992)128. }
\lref \frts {E.S. Fradkin and A.A. Tseytlin, \np B261(1985)1. }
\lref \mplt {A.A. Tseytlin, \mpl A6(1991)1721. }
\lref\wittt { E. Witten, \cmp 121(1989)351; G. Moore and N. Seiberg, \pl
   B220(1989)422.}
\lref \chernsim { E. Guadagnini, M. Martellini and M.
   Mintchev, \np B330(1990)575;
   L. Alvarez-Gaume, J. Labastida and A. Ramallo, \np B354(1990)103;
   G. Giavarini, C.P. Martin and F. Ruiz Ruiz, \np B381(1992)222; preprint
   LPTHE-92-42.}
\lref \polles { A. Polyakov, Les Houches Lectures (1988).   }
\lref \kutas { D. Kutasov, \pl B233(1989)369.}
\lref \kniz {  V. Knizhnik and A. Zamolodchikov, \np B247(1984)83. }
\lref \anton { I. Antoniadis, C. Bachas, J. Ellis and D.V. Nanopoulos, \pl B211
   (1988)393.}
\lref \ts  {A.A.Tseytlin, \pl B268(1991)175. }
\lref \shts {A.S. Schwarz and A.A. Tseytlin,  preprint Imperial/TP/92-93/01. }
\lref \bush { T.H. Buscher, \pl B201(1988)466.  }
\lref \plwave { D. Amati and C. Klim\v cik, \pl B219(1989)443; G. Horowitz and
   A. Steif, \prl 64(1990)260.}
\lref \div  { P. Di Vecchia,  V. Knizhnik, J. Peterson and P. Rossi, \np
   B253(1985)701. }
\lref\IBvertex{I. Bars, in {\it Vertex Operators in Math and Physics}, Eds.
   Lepowsky, Mandelstam, Singer, (Springer Verlag NY, 1985), page 373.}
\lref\grt{A. Giveon, E. Rabinovici, and A.A. Tseytlin, Heterotic String
Solutions and Coset Conformal Field Theories, hep-th/ 9304155}


 hep-th/9309042        \hfill       USC-93/HEP-B3
  \bigskip

\centerline {\bf CURVED  SPACETIME GEOMETRY FOR STRINGS }
\centerline { {\bf AND  AFFINE NON-COMPACT ALGEBRAS }
  \foot{Based on lectures delivered at the  (1) AMS meeting at USC, Nov. 1992,
(2) Conference on Quantum Aspects of Black Holes, ITP, UC-Santa Barbara, June
1993, and
\hb (3) 25$^{th}$ Summer Institute, \'Ecole Normale Sup\'erieure, Paris,
France, Aug. 1993.
 \hb To appear in {\it Interface Between Mathematics and Physics}, Ed. S.-T.
Yau .   }  }

 \bigskip\bigskip
\centerline{ { ITZHAK BARS }
 \foot{Research partially supported by DOE grant No. DE-FG03-84ER-40168 and
by NSF Grant No. PHY89-04035 .} }

 \bigskip
\centerline{ { Department of Physics and Astronomy } }
\centerline{  { University of Southern California}  }
\centerline{  { Los Angeles, CA 90089-0484, USA}   }

\bigskip\bigskip

\centerline {ABSTRACT}
\bigskip

\item{1--} Itroduction to String Theory in Curved Spacetime
\item{2--} G/H Coset Conformal Field Theory and String Theory

\item{3--} Time, Space and Classification of Non-Compact Cosets

\item{4--} Heterotic Strings in Curved Spacetime as gauged WZW Models

\item{5--} The Spacetime Manifold and the Geometry

\item{6--} Examples in 2D, 3D and 4D

\bigskip

\overfullrule=0pt

\bigskip

\newsec { INTRODUCTION }

This introduction  will give a brief summary of the physical ideas
and mathematical concepts without going into the details of the physical
models. Its purpose is to introduce some definitions and provide some
guidance for the non-expert mathematician or physicist.
In the first part of the introduction the physical motivation and
scenario will be introduced, while in the second part the formalism and
models
will be summarized. If some terms
seem unfamiliar, the reader should not be discouraged, because either
the jargon used is not important or it will be defined at some later
point. In the following sections more details will be given. The
mathematically inclined reader who may not care about the physical
applications may want to skip the first part of this introduction.

\subsec{Physical Motivation and Scenario}

I first want to make a case for studying String Theory in
curved spacetime. I will mention quantum gravity and the unification
of forces as the two main reasons.

It is generally agreed that quantum gravity is an obscure frontier.
Every discussion of black holes or similar singularities fizzles out
at the vicinity of the singularity and tries to put the blame on
the lack of understanding of quantum gravity. One of the main reasons
for this is that ordinary gravity is not a renormalizable quantum
field theory. On the other hand, String Theory is the main known
candidate for a finite theory of quantum gravity. However, in String
Theory most discussions of gravity are perturbative, and in this form
one cannot address issues of singularities and strong gravitational
fields. Therefore, String Theory in curved spacetime must be studied
in order to understand how string theory handles the issues of black
holes and other singularities. Furthermore, it is important to tackle
the problems of quantum gravity in a complete theory such as String
Theory, since otherwise some important features may be missing in the
discussion. For example, spacetimes with duality and mirror symmetry
properties emerge in String Theory. It is not known whether such
features are important for quantum gravity, but they are entirely
missing in the traditional discussions of quantum gravity.

One of the main promises of String Theory is
to explain the common origin of all forces
and all matter.
In the mainstream of String Theory this unification has been
formulated by assuming a four dimensional {\it flat spacetime}
plus additional {\it curved compactified} spacelike
dimensions described by a conformal field theory, with the condition
that the total theory is conformally critical. This cannot be
the full story, since in this form gravity is perturbative,
and one cannot discuss the effects of strong gravitational fields
or singularities (such as the Big Bang) on the formation
of matter during the early part of the Universe. Therefore,
predictions on the unification of forces and the nature of matter
such as the quark
 and lepton families may be incomplete or even
wrong.

Therefore, I have proposed the following scenario in which I
believe String Theory plays a role. I assume that during the
evolution of the Universe there is a String Era during which
spacetime is curved. There has been speculation that at the
earliest times spacetime dissolves, there is no metric, and
only topological notions survive (Witten). If this is true, at
slightly later times a String Era must begin, and my discussion
of String Theory would apply at those times. Once one accepts
the existence of a String Era, described by a string propagating
in a curved spacetime, then the mathematical consistency of the
theory (i.e. conformal invariance at the quantum level) does
not necessarily require the existence of higher dimensions.
Maybe there are extra dimensions beyond four, maybe not. It then
becomes attractive to speculate that, in fact there are only
four dimensions at all times, i.e. during the String Era and
afterwards. (My discussion below is not necessarily committed to
only four dimensions and it will apply to any number of dimensions
from two to 26. In fact, it is perhaps embarrassing that the number
of dimensions during the String Era is mathematically allowed to be
less than, equal to or larger than four). I assume that during the String
Era the Universe is described by a heterotic string theory (of
the type given below) which provides a complete theory. It is during
this era that the gauge symmetries and the matter content, i.e.
quark and lepton families, are determined. When the theory makes
a transition to the eventual flat spacetime of our current Universe
the symmetry and matter content is expected to survive.
In particular, I point out that, even in curved spacetime one can
identify the low energy matter as being the lowest string states
classified in {\it chiral gauge multiplets}, since these will
be protected from getting a mass in either flat or curved spacetime.

Therefore,
I believe that, in order to realize the prediction of String Theory
for the unification of forces and matter, one must study string
theory in curved spacetime during a String Era. If I am correct
about this point, the earlier studies with a flat four dimensional
spacetime, plus Calabi-Yau manifolds (etc.), may have been on the wrong
track. As will be seen in examples given below, the gauge
symmetries that emerge in {\it purely four dimensional} heterotic
superstring models (in curved spacetime)
 contain the Standard Model or Grand Unified
Model symmetries, such as SU(3)xSU(2)xU(1) or SU(5) or SO(10). In
this sense there is already an encouraging sign that the
curved spacetime approach is promising.

There are also some obvious problems that need more understanding
than anyone can offer at the present. In particular
let us consider
the transition from the curved spacetime to the flat four
dimensional Universe.
I will assume that several phase transitions
that must occur in any string theory will help make this
transition. First, one knows that there has to be a phase
transition that gives the dilaton a mass. Second, there has to
be a phase transition that inflates a small part of the early
universe to the known part of the current Universe. These
phase transitions must occur even if string theory is formulated
by starting with a flat four dimensional spacetime
or with a curved spacetime in more than four dimensions.
It is expected that the gauge forces drive these phase
transitions. Such gauge forces are present in a heterotic
string theory. Unfortunately,
these phase transitions in string theory have not yet been understood,
and I have nothing new to add to this at this time. However, I
emphasize that because of such expected phase transitions, it
is perfectly plausible that even though we start from a curved
spacetime that is inhomogeneous and non-isotropic, the inflated
current Universe that is flat can emerge from any part of the
universe of the String Era. In particular, as will be seen below,
the String Era geometry has an asymptotically flat region that is
described by the linear dilaton background.  The final flat Universe
may be the inflation of a small region of this asymptotic geometry.
If there are more dimensions than four, the linear dilaton
may point along the extra dimensions. If there are only four
dimensions a linear dilaton would violate Poincar\'e invariance.
However, because of inflation, the dilaton will take its values in
only a small region of the initial
 space, thus appearing as almost a constant
in the inflated universe, and therefore consistent with
an apparent Poincar\'e
invariance. In describing the final flat Universe one must use
an effective low energy action {\it after all the phase transitions}
have taken place. The part that remains obscure in this scenario
is how to derive this final effective action and verify that
indeed our kind of universe emerges.

The remarks above summarize my  ``religion" on the subject
of unification.
Having explained my point of view, I can go on with the technical
aspects which may have applications whether or not the universe is
purely four dimensional or not during the String Era. In order to
study string theory in curved spacetime I have formulated exactly
solvable models in which many questions can be more easily investigated.
These models are based on non-compact Kac-Moody algebras which
provide a Hamiltonian formulation of gauged Wess-Zumino-Witten models.
The non-compact groups and an appropriate set of their cosets are
chosen to describe spacetime with {\it a single time coordinate}.
I now give a brief description of these models.

\subsec{Formalism and models }
More generally, there
are models in string theory that are based on Wess-Zumino-Witten
(WZW) models, or {\it gauged} Wess-Zumino-Witten models
(GWZW), in which the quantum commutation rules \WITT\IBvertex\
can be put into the form of an affine  Kac-Moody or
Bardakci-Halpern algebra \KAC .  One starts with a map from a Riemann
surface (e.g. the sphere, parametrized by the
complex coordinate $z$ ) to a group G. The map $g(z,\bar z)$ is a matrix
in a representation of G, usually taken as the fundamental representation.
Then one constructs the
``currents" that are elements in the left Lie algebra,
$it^AJ_A(z)=g^{-1}\partial_zg$, and the right Lie
algebra, $it^A\bar J_A(\bar z)=\partial_{\bar z}gg^{-1}$.
These are holomorphic and anti-holomorphic respectively \WITT .
The matrices $t^A, A=1,2\cdots ,dim\ G$,  represent the
Lie algebra (whose dimension is $dim\ G$) and they provide a basis for it.
One can define string coordinates $X_A(z,\bar z)$ which provide
a parametrization of the group manifold $g=g(X(z,\bar z))$, for example
$g(X)=exp(it^AX_A)$ (later other more convenient parametrizations
will be used). Then the currents are really constructed from the
string coordinates $J_A=\partial_zX_A+\cdots$ and $\bar
J_A=\partial_{\bar z}X_A+\cdots$, where  the dots
$\cdots$ are non-linear terms
in $X_A$.

One expands these currents as a Laurent series in powers of $z$
(e.g. on the sphere) and identifies the expansion coefficients as
the generators of the affine algebra as follows: left currents
$J_A(z)=\sum_n J_{An} z^{-1-n}\quad $, right currents $J_A(\bar
z)=\sum_n \bar J_{An} \bar z^{-1-n}$. The canonical commutation
rules that follow from the standard quantum theory take
the form of the affine algebra

\eqn\affine{ [J_{An},J_{Bm}]=if_{AB}\ ^C J_{C,n+m} -k{n\over
2}\eta_{AB}\delta_{n+m,0} }
where, in an appropriate basis, the Killing form is diagonal
$\eta_{AB}=diag (++\cdots, --\cdots)$, with $+1$ entries for
compact generators and $-1$ entries for non-compact ones. The
commutation rules look identical for the right moving
(anti-holomorphic) currents $\bar J_{An}$.
The constant $k$ is called the central extension, and is a parameter
in the WZW action \WITT\IBvertex .

There is a quadratic form in the currents  (called the
Sugawara form) which has a structure similar to the quadratic Casimir operator
(but is not an invariant). Actually, because of the infinite
dimensional nature of the algebra there are an infinite number of
such quadratic forms which, taken together, close among themselves
under commutation and, form the Virasoro algebra.
These are given by $L^G_n=\sum_m :J_{A,-m}J_{B,m+n}:\eta^{AB}/(-k+g)$,
where $g$ is the Coxeter number for the group $G$, and the colons  ``:"
indicate  ``normal ordering". The stress tensor of the WZW model
has Fourier coefficients that are precisely these Virasoro generators.
The Virasoro generators are the generators of infinitesimal conformal
transformations which play a fundamental role in string theory.
The sum of the zero modes of the holomorphic and anti-holomorphic
sectors, $L^G_0+\bar L^G_0$, is the Hamiltonian of the WZW model.
When expressed in terms of the string coordinates $X^A$ introduced above
this Hamiltonian is rather non-linear. As we shall see later
the structure of these operators in terms of the string coordinates
is what defines {\it the geometry of spacetime.}

The so called coset models \GKO\ of string theory correspond to {\it
gauged} WZW models and they work as follows. Consider an infinite
dimensional affine algebra for the group G, as in \affine .  Consider a
subgroup $H\subset G$ and its corresponding affine algebra denoted by
$J_{an}, \ a=1,2,\cdots ,dimH, \ n\in \Z$. This subgroup is gauged, and
the gauge currents are the $J_{an}$. The stress tensor for the
$G/H$ coset model for string theory is constructed by subtracting the
quadratic Sugawara form for $H$ from the quadratic Sugawara form for
$G$. This gives a new Virasoro algebra $L_n=L^G_n-L^H_n$
\foot{The $L^H_n=\sum_m:J_{a,-m}J_{b,m+n}:\eta^{AB}/(-k+h)$ have a structure
similar to $L^G_n$ above, where the metric
$\eta^{ab}$ and Coxeter number $h$ are appropriate to the
subgroup $H$.}. An important
feature of the $L_n$ is that they are gauge invariant, i.e. they
commute with the gauge currents of the subgroup, $[L_m, J_{an}]=0$. The
$G/H$ coset scheme is very tight because of a requirement of
conformal invariance that underlies string theory.  Conformal
invariance is successfully incorporated by demanding the closure of
the Virasoro algebra mentioned above, and this is achieved by the
scheme automatically. This feature makes the $G/H$ model an exactly
solvable string theory by using the representation theory of the
affine algebra.

Modules of the affine algebra are used to construct the
Hilbert space of the string theory. One considers the direct
product of left-mover (holomorphic) and right-mover
(anti-holomorphic) sectors. The string theory Hilbert space is
obtained as  ``modular invariant" combinations of the left and right
modules (modular invariance will not be explained here, however, it is
guaranteed by a Lagrangian. The exact string models presented in this paper
have such a Lagrangian construction in the form of gauged WZW models.).
Furthermore,
the string theory comes with constraints on the canonical variables.
The physical states of the theory are identified as the subset of
states in the Hilbert space on which the constraints vanish. One set
of constraints is that currents that belong to the subalgebra of $H\subset
G$ must vanish: $J_{an}=\bar J_{an}=0$ (on kets) for $n\ge=1,\
$ . This is equivalent to demanding gauge invariant
physical states. Another set of constraints is that the generators of
the Virasoro algebras in the holomorphic and the anti-holomorphic
sectors vanish, $L_n=\bar L_n=0$ (on kets) for $n\ge 1$. This is equivalent
to requiring reparametrization invariant physical states. The zero modes
$J^G_{A0}, L_0$ do not vanish on physical states (we will append the
extra letter G or H to the currents in order to emphasize that
they belong to the Lie algebra of the group G or the subgroup H) .

Concentrate on the ground state of the string theory, which
is called the Tachyon state T. It satisfies the conditions

\eqn\tachy{ (L_0+\bar L_0-2)T= (J^H_0 +\bar J^H_0 )T=0, \qquad
J^G_n T= \bar J^G_n T =0, \quad n\ge 1}
Only the zero modes $J^G_{A0} ,\bar J^G_{A0}$ are non-zero on the Tachyon.
Therefore, only the zero mode of the group element $g\in G$ is
relevant for this state $T(g)$, and one can ignore the Riemann surface
$(z,\bar z)$. Note that for the subgroup $H$ the combination
$J_{a0}+\bar J_{a0}$ must vanish.  This last requirement is also a
remnant of gauge invariance. It demands that the tachyon is a singlet
under the action of the subgroup $H$ when it acts according to the
adjoint action, i.e.

\eqn\adj {T(hgh^{-1})=T(g) \ .}
Therefore the tachyon is a function of the coset $G/H$, where $H$
acts according to the {\it adjoint action}. This coset defines {\it
the manifold for our geometry}. The spacetime geometry on which the
string propagates is identical to the geometry that arises through
the tachyon state (this point will not be further explained here).

To find the metric and other properties of the geometry
we analyze further the eigenvalue equation in \tachy\ that involves
the Hamiltonian = $L_0+\bar L_0$, constructed from the $G/H$
coset Virasoro operators:

\eqn\cas { L_0=-{J_G\cdot J_G\over k-g}-{J_H\cdot J_H\over k-h}, \qquad
\bar L_0=-{\bar J_G\cdot \bar J_G\over k-g}-{\bar J_H\cdot \bar
J_H\over k- h}\ . }
Since only the zero modes $J^G_{A0}, \bar J^G_{A0}$ enter, these are
just a combination of the quadratic Casimir operators of $G$ and of $H$.
By writing these Casimir operators in the form of second order
differential operators in the string variables $X^\mu$ (see below)  we
define the metric $G_{\mu\nu}(X)$ and the dilaton $\Phi (X)$

\eqn\laplacian{ (L_0+\bar L_0)T={-1\over e^\Phi\sqrt{-
G}}\partial_\mu(e^\Phi\sqrt{-G}G^{\mu\nu}\partial_\nu T) \ .}
By working out the left
hand side of this equation through group theoretical manipulations
and then comparing to the right hand side
 we can read off the metric defined on the manifold
$G/H$.  This metric and dilaton automatically solve the Einstein
equations for dilaton gravity in the large $k$ limit. The Einstein
equations impose conformal invariance perturbatively in lowest order
in an expansion of the theory in powers of $k$. However, at finite
$k$ our expressions give the conformally exact metric and dilaton.

The spectrum is given by the unitary representations of $G$
since all that is needed is to diagonalize the quadratic
Casimir operators of the left and right groups in a basis
that is {\it restricted to H} .  Therefore, $T(g)$
must be a linear combination of the D-functions for the group in the
unitary representation R of $G$, with a trace over the subgroup $H$  in
any of the representations of $H$ (to impose a singlet of $H$).
Symbolically

\eqn\tachysol{ T(g)= Tr_H(D^R(g)) \ .}
Only the representations $R$  and its restrictions to $H$
that yield the eigenvalue $L_0+\bar L_0=2$
are admitted as solutions.

Three years ago it was understood that, by appropriately
choosing $G$ and $H$, one can describe strings propagating in gravitational
backgrounds that correspond to curved spacetime with a single time coordinate
 \BN . All possible cosets that have the single time
property have been found and classified
\IBCS\GIN . These have also been extended by supersymmetry and
classified \IBhetero .
 Furthermore, by using the equivalence of the $G/H$ scheme to the gauged
Wess-Zumino-Witten model (GWZW) it has been possible to connect the
geometry of spacetime to the algebraic scheme \BN\ \WIT .  The first
example $SL(2,\IR)/SO(1,1)$ which was worked out explicitly was
interpreted as a string propagating in the gravitational background
of a black hole in two dimensions \WIT . By now there are examples in
three and four dimensions as well as with supersymmetry. These
geometries solve the Einstein equations in dilaton gravity and have
singularities that are more intricate than black holes. This kind
of gravitational singularities have not arisen and have not been studied
before. Furthermore, the global manifold turns out to have certain
``duality" properties, by which we mean that there is a symmetry
transformation that does not change the spectrum but interchanges
patches of the manifold. Such manifolds and properties are of great
interest for quantum gravity as well as string theory as applied to the
Early Universe or to black hole physics. The fact that the models
outlined above are completely solvable make them very attractive for
studying the fundamental questions that arise in these physical
phenomena.

As seen from the above summary, this approach involves several
fields of mathematics (algebra, geometry, representation theory,
differential equations, analysis) on the one hand and string theory
and conformal field theory on the other. The fact that the groups $G$ of
interest are non-compact means that little is known about these
models at the present, and much more needs to be done.

In the remainder of this paper we will provide more details by giving
a summary of the coset scheme in section 2, while examples of the singular
gravitational geometries and some of their properties will be
outlined in section 3.

\newsec {THE COSET SCHEME IN STRING THEORY}

\subsec{ An introduction to affine algebras and WZW models}

Let us consider an affine algebra associated with a group G which may be
compact or non-compact \KAC

 \eqn\KM{ [ J_{nA}, J_{mB}] =
if_{AB}^C  J_{n+m,C} -n{ k \over 2}
    \eta_{AB} \delta_{n+m},    }
where $\eta_{AB}$ is taken proportional to the Killing metric and is defined by
$ g_{AB}= \ f_{AC}^D f_{BD}^C = \ g \eta_{AB}. $ The number $g$ is the Coxeter
number, e.g. for $SU(N)$, $g=N$; $k$ is positive and larger than $g$
when $G$ is non-compact and a negative {\it integer} when $G$ is
compact. Furthermore, in the non-compact case, $k$ must be a positive
integer larger than $g$ if the maximal compact subgroup of  $G$ is
non-abelian. The generators of the affine algebra are Fourier
components of a local current defined on the string worldsheet
$J_A(z)=\sum_{n=-\infty}^{+\infty}J_{nA}z^{- n-1}$
($z=exp(\tau+i\sigma)$ and $(\tau ,\sigma)$ parametrize the Euclidean
continuation of the worldsheet). It may be constructed from a group
element $g(z,\bar z)\in G$ as $t^AJ_A=-ig^{-1}\partial_z g$, where
$t^A$ is a basis of matrices for the Lie algebra, say in the
fundamental representation. After writing an appropriate action,
which is the Wess-Zumino-Witten (WZW) action, and quantizing it, one
finds that the commutation rules above correspond to the canonical
quantization of the model \WITT\IBvertex .  Actually one finds that
there is another current, $t^A\bar J_A(\bar z)=-i\partial_{\bar z}g
g^{-1}$, whose Fourier components $\bar J_{nA}$ are canonical degrees of
freedom independent from $J_{nA}$, and form another affine
algebra with the same structure as above. The WZW action that gives
rise to these structures may be written as (in the Minkowski version
of the worldsheet, with $\sigma^\pm =(\tau \pm \sigma)/\sqrt{2}$):

 \eqn\action{
 S_0(g)={k\over 8\pi}\int_M d^2\sigma\ Tr(g^{-1}\partial_+g\ g^{-1}\partial_-g)
 -{k\over 24\pi}\int_B Tr(g^{-1}dg\ g^{-1}dg\ g^{-1}dg) . }

In order to gain a bit more insight into the meaning of these equations
it is useful to consider the large $k$ limit. Large $k$ is equivalent to small
$\hbar$ and hence it corresponds to the semi-classical limit of the
theory (since the path integral involves the action in the form
$exp(iS/\hbar)$).  In this limit it is useful to rescale the
generators by defining $\alpha_{nA}=J_{nA}/\sqrt{k}$ and then taking
the $k\to \infty$ limit in eq.\KM . The result is

\eqn\osc{[\alpha_{nA},\alpha_{mB}]=-n\eta_{AB}\delta_{n+m,0}} %
 which
is equivalent to harmonic oscillator commutation rules. A similar
result holds for $\bar\alpha_{nA}=\bar J_{nA}/\sqrt{k}$. Thus, the
generators of the affine algebra may be thought of as a  non-Abelian
version of harmonic oscillators. This can be made more explicit as
follows.  With an appropriate parametrization of the group element by
coordinates $X_A$, say $g=exp(it^AX_A/\sqrt{k})$, the large $k$ limit
of the action reduces to an action for the free field $X_A(z,\bar
z)$. The $X_A(z,\bar z)$ will be interpreted as the string
coordinates defined on the Euclidean worldsheet $(z,\bar z)$, with
$z=exp(\tau+i\sigma)$. In fact, in the large $k$ limit, the free
field can be thought of as a free string propagating in flat
spacetime.  For finite $k$ the spacetime will be curved. At this
point the signature of spacetime is not yet correct (i.e. one needs
only one time coordinate and additional space coordinates), but we
will see later how to construct a good model. The Fourier components
of the free field are, in fact, the $\alpha_{nA}, \bar\alpha_{nA}$
oscillators. The perturbative expansion of the non-linear theory in
powers of $1/k$ can be worked out by starting with the free field
Hilbert space defined by the Fock space of the
$\alpha_{nA},\bar\alpha_{nA}$. This will correspond to an expansion
around flat spacetime. However, the theory can also be solved
non-perturbatively for any $k$, in curved spacetime, by using the
representation theory of the affine algebra, and this is the
principal advantage of considering such a theory.

The conserved energy-momentum tensor has components $T^G_{zz},T^G_{\bar z\bar
z},T^G_{z\bar z}$ (using the letter $G$ to emphasize that the group
$G$ is involved). After taking quantum ordering problems into account
they take the form $T^G_{zz}=:J_A(z)J_B(z):\eta^{AB}/(-k+g)$ with a
similar form for $T^G_{\bar z \bar z}$ in terms of $\bar J_A$, while
$T^G_{z\bar z}$ vanishes.  The colon ``:" is used to indicate normal
ordering. Note that the form of $T^G$ has the same structure of the
quadratic Casimir operator, and it is called the Sugawara form. It
has the Fourier expansion $T^G_{zz}=\sum_n L^G_n z^{-n-2}$, where the
$L^G_n$ are given in the introduction. The
$L^G_n$ are called Virasoro generators and they could be considered
to be a kind of generalization of the quadratic Casimir operator.
However, unlike the usual Casimir operator, the Virasoro operators do
not commute with the currents or with each other, rather

\eqn\vir{ [L^G_n,L^G_m]=(n-m) L^G_{n+m} + {c_G\over 12} n(n^2-
1)\delta_{n+m,0},
          \qquad [L_n^G,J_{mA}]=-m J_{n+m,A} .}
where the so called Virasoro central charge is $c_G=k\ dim(G)/(k-g)$.
Similarly one constructs $\bar L^G_n$ from the $\bar J^G_{An}$ with
identical properties. The
Hamiltonian of the theory is constructed from the zero mode Virasoro
operators and is given by $H=L^G_0+\bar L^G_0$.

The eigenstates of the Hamiltonian are formed by constructing the
representations of the affine algebra as follows. First one defines the
so called ground states, or level zero states, $|R>_G^0$, which correspond to
the states in any unitary representation $R$ of the group G. One assumes that
these states form the vacuum states for the ``non-Abelian oscillators". For
$n\ge 1$ the $J_{-n,A},J_{n,A}$ are considered creators and annihilators
respectively, while for $n=0$ the $J_{0,A}$ act via a matrix representation
$t^R_A$ appropriate for the representation $R$

\eqn\annih{ J_{0,A}|R>_G^0= t^R_A|R>^0_G ,\qquad J_{nA}|R>^0_G=0,\quad n\ge
1\ .}
$|R>^0_G$ symbolizes the basis states in the representation $R$ of
the ordinary Lie algebra of $G$. The excited higher level states are
obtained by applying integer powers $p_i$ of the creators
$J_{-n_i,A_i},\ n_i\ge 1$, on the ground state, thus constructing the
Hilbert space

 \eqn\states{ |R>^l_G=\prod_i\bigl (J_{-n_i,A_i})^{p_i}\bigr )|R>^0_G\ .}
The level is given by $l=\sum_i n_ip_i$. This space is somewhat analogous to
Fock space. In fact, for large $k$ the states \states\ reduce to ordinary Fock
space constructed from the $\alpha_{nA}$ oscillators introduced above. Of
course, the non-Abelian nature of the creators provide for a richer structure.
The eigenvalue of the Hamiltonian is then obtained from

 \eqn\eigen{ L_0^G|R>^l_G=({C(R)\over {-k+g}} +l)|R>_G^l  \ .}
where $C(R)$ is the eigenvalue of the usual quadratic Casimir operator for
representation $R$. A similar result is obtained for $\bar L^G_0$. Thus, for
any value of $k$ {\it the spectrum is calculated from the quadratic Casimir. }
Furthermore, the non-linearity of the theory has been shoved entirely into the
Casimir operator through the representation content $R$ of the state. The
excitations contribute to the energy through the level, the integer $l$, in a
way that is quite analogous to excitations in a Fock space.  As we shall see,
the geometry of spacetime is also embedded in the Casimir operator.

\subsec {Introduction to Cosets and Gauged WZW models}

The action \action\ may be modified by gauging a subgroup $H$ of $G$. The new
action takes the form $S=S_0+S_1$ with \GWZW

\eqn\actionone{S_1(g,A)=-{k\over 4\pi}\int_M d^2\sigma\ Tr(A_-\partial_+gg^{-1}
-\tilde A_+g^{-1}\partial_-g + A_-g \tilde A_+g^{-1}-A_-A_+) \ .}
where the gauge fields $A_+,A_-$ are matrices in the Lie algebra of $H$. (A
generalization of this action with $A_+$ twisted relative to $A_-$ exists
\BSthree , and this is important for a phenomenon called ``duality", but we
will not be concerned with it in this paper.) The gauge invariance leads to
constraints that require the subgroup currents $J_{n,i}, \bar J_{n,i}$, with
$i\in H$, to vanish. In the quantum theory the constraints are applied on the
states by demanding the subgroup currents with $n\ge 1$ to vanish on kets
$|R>_G^l$, and those with $n\le 1$ to vanish on bras, while for $n=0$ one must
take only the combination $J_{0,i}+\bar J_{0,i}$ to vanish. The physical
states are only those states that satisfy the constraint equations.

With these constraints one is effectively removing gauge degrees of freedom
that correspond to the subgroup $H$, thus dealing with a theory whose degrees
of freedom are in some sense associated with the coset\ $G/H$. Again, it is
useful to consider the large $k$ limit, for which the constraints are solved
very simply: Consider the oscillators
$\alpha_{An}=(\alpha_{an},\alpha_{\mu n})$, where the first set is
assiciated with $H$ and the second set with $G/H$, and take only the
Fock space constructed from the oscillators $\alpha_{n,\mu},\bar
\alpha_{n,\mu}$ with $\mu\in G/H$ and ignore the subgroup oscillators
(this is equivalent to a free string $X_\mu (\tau,\sigma)$ in flat
spacetime). The removal of degrees of freedom is a consequence of the
subgroup gauge invariance: one may imagine choosing a gauge in which
the $H$-gauge fields ``eat up" degrees of freedom from the group
element $g\in G$. The gauge fields are not dynamical since they
appear without derivatives; thus they are just Lagrange multipliers
which, through the equations of motion, are functions of the remaining
degrees of freedom. Therefore the gauged WZW model indeed depends
only $G/H$ degrees of freedom. For finite $k$ the explicit dependence
of the theory on the $G/H$ degrees of freedom is highly non-trivial
and is different than the old $G/H$ sigma models. Furthermore, it
contains the information on the geometry, as will be seen later.

The energy-momentum tensor for the gauged WZW model is constructed by
subracting the Sugawara form for the subgroup $H$ from the corresponding form
for $G$ \GKO

\eqn\stresscos{ T^{G/H}_{z\bar z}= {:J_AJ_B:\eta^{AB}\over -k+g}-
             {:J_iJ_j:\eta^{ij}\over -k+h}\ , }
where $g,h$ are the Coxeter numbers for $G,H$ respectively.
 Therefore the Virasoro generators for the gauged WZW model are given by
$L^{G/H}_n=L_n^G-L_n^H$, and therefore the Hamiltonian is $H=L_0^{G/H}+\bar
L_0^{G/H}$.

The eigenvalues of this Hamiltonian are computed as follows: First
decompose the representation $R$ into representations $r_h$ of the subgroup
$H$, i.e. $|R>_G^l=\sum_h \oplus |r_h>_H^{l_h}$, then compute the difference
between the Casimir eigenvalues

\eqn\eigen{ L_0^{G/H}|r_h>_H^{l_h}=\bigl [{C^G(R)\over -k+g}-{C^H(r_h)\over -
k+h}+(l-l_h) \bigr ] |r_h>_H^{l_h} \ . }

In the full string theory that we will consider below we also need to impose
the gauge constraints that follows from the reparametrization invariance, or
conformal invariance of the string theory. This corresponds to the vanishing
of the Virasoro generators $L^{G/H}_n, \bar L_n^{G/H}$, for $n\ge 1$, on
physical kets, and demanding $L_0^{G/H}=\bar L_0^{G/H}$ as well.
Therefore, the eigenvalue of the Hamiltonian on a physical state is
$H=2L_0^{G/H}$. A physical state involves both sectors of the Hilbert
space constructed from the $J_{nA}$ and $\bar J_{nA}$ in ``modular
invariant" combinations (not to be explained here).

These
constraints, as well as the $G/H$ constraints outlined above, are all
automatically satisfied for the ground states ($l=0=l_h$) that are
gauge invariant under the subgroup $H$. These are called the
tachyonic states (although they are not necessarily tachyons in a
supersymmetric theory, or in curved spacetime). Thus, a physical
state of the gauged WZW model at the tachyonic level (i.e. ground
state level) is automatically obtained for any state that satisfies

\eqn\tach{ (J_{0,i}+\bar J_{0,i})|Tachyon, R, r_h>=0 . }
This is nothing but the requirement of gauge invariance with respect
to the subgroup $H$. The energy
eigenvalue follows from \eigen . We will examine the states that satisfy these
conditions in order to extract the spacetime geometry.

\subsec{Time Coordinate}

Let me review how the single time coordinate condition restricts the possible
cosets $G/H$. Let us consider a WZW model based on a non-compact group.
Let us parametrize the group element by $X^A(\tau,\sigma)$, where $A$ is an
index in the adjoint representation. The left or right moving currents take
the form $J^A=\partial X^A+\cdots$, where the dots stand for non-linear terms
in an expansion in powers of $X$. The Fourier components of these currents
$J_n^A$ satisfy an affine algebra as in eq.\KM\ ,
where $k$ is the central extension and $\eta^{AB}$ is proportional to the
Killing metric. In an appropriate basis one can choose a diagonal $\eta^{AB}=
diag (1,\cdots,1, -1,\cdots -1)$ with $+1$ entries corresponding to compact
generators and $-1$ entries to non-compact ones. For example, for $SL(2,\IR)$
with currents $(J^0,J^1,J^2)$, one has the Minkowski metric in $2+1$
dimensions: $\eta^{AB}=diag (1,-1,-1)$.

As explained above, when $k\to\infty$ the currents behave like the free field
oscillators of the flat string theory as in \osc . Examining the signature of
the oscillators as given by the Killing metric (with the extra sign in front)
we interpret the free fields $X^A\sim \sum_n {1\over n}\alpha_n^A z^n +
\cdots$ as time coordinates when $A$ corresponds to compact generators and as
space coordinates when $A$ corresponds to  non-compact generators. The
signature of the coordinates are the same for finite positive $k$. This is
seen by specializing the commutation rules \KM\ to $A=B$ for which the
structure constant of the Lie algebra drops out. The timelike oscillators
create negative norm states that ruin the unitarity of the theory. The
negative norm states must be eliminated from the theory by a consistent set of
constraints.

For ordinary string theory in flat spacetime the necessary
constraint emerges automatically from the conformal invariance of the theory.
Conformal invariance requires the vanishing of the Virasoro generators, and it
can be shown that indeed these conditions remove all negative norms. This
result goes under the name of the no-ghost theorem.

In a string theory one can tolerate only one time coordinate. This is because,
by naive counting, the Virasoro constraints $L^{G/H}_n\sim 0$ can eliminate
only the ghosts generated by the negative norm of one time-like oscillator
$\alpha_n^0$, just like string theory in flat spacetime. Therefore, one must
put constraints that set to zero the unwanted time-like compact generators
$J_{n,i}$, except for one of them. However, first class constraints of this
type must close to form an algebra. Therefore, the currents that are set equal
to zero ($J^i\sim 0$ weakly on states) must form a subalgebra corresponding to
a subgroup of the non-compact group $H\subset G$. The subalgebra may include
compact and non-compact generators. The remaining currents $J^\mu, \
\mu=0,1,2,\cdots (d-1)$ stand in one-to-one correspondance with the coset
coordinates $X^\mu$ that include just one time coordinate. Thus, one must
choose a subgroup $H$ such that the coset $G/H$ has the signature of Minkowski
space $\eta_{\mu\nu}$ in $d$ dimensions. As seen in the previous section this
set of constraints defines an exact conformal field theory that fits the
algebraic framework of the gauged WZW model. The important ingredient is that
one must take an appropriate non-compact coset $G/H$. The only simple groups
that give a single time coordinate were classified in \IBCS

\eqn\list{\eqalign {
  SO(d-1,2)/SO(d-1,1) \qquad & SO(d,1)/SO(d-1,1) \cr
  SU(n,m)/SU(n)\times SU(m) \qquad & SO(n,2)/SO(n) \cr
  SO(2n)^*/SU(n)        \qquad &  Sp(2n)^*/SU(n)    \cr
  E_6^*/SO(10)         \qquad  &  E_7^*/E_6 } }
This list, which contains only simple groups, may be extended with direct
products of simple groups $G_1\times G_2\times \cdots$ including $U(1)$ or
$\IR$ factors, or their cosets, so long as the additional factors do not
introduce additional time coordinates \BN\IBCS\GIN .

There is another way to see the same result by using a Lagrangian method at
the classical level rather than the algebraic Hamiltonian argument given above
at the quantum level. As discussed in the previous section, a coset theory
corresponds to a gauged WZW model with the
subgroup $H$ local. Using the gauge invariance one can eat up $dim(H)$ degrees
of freedom, leaving behind $dim(G/H)$ group parameters that contain just one
timelike coordinate. Since the gauge fields are non-dynamical they can be
integrated out. This leaves behind a sigma model type theory with the desired
signature. The large $k$ limit of this theory has free field quantum
oscillators with a single time coordinate.

Both the Hamiltonian and Lagrangian arguments were first given by Bars and
Nemeschansky \BN . The Hamiltonian approach was given more weight in \BN\
where several examples, including $SL(2,\IR)/\IR$ at $k=9/4$, were
investigated. It was found that this model describes a black hole in two
dimensions \WIT\
\foot{This model, although first suggested in \BN\ as a candidate for strings
in curved spacetimes, got popularly known as the Witten Black Hole \WIT , since
he carried out the Lagrangian argument explicitly and interpreted the sigma
model metric as a black hole. It should not be very difficult to find a more
appropriate name that recognizes the original discovery.}.
 With the realization
that non-compact group coset methods generate singular geometries, there has
been a flurry of activity to determine the geometries of higher dimensional
cosets \BSthree\others . While these models represent only a small subset of
all possible curved spacetime models described by the general sigma model,
they have the advantage of being solvable in principle thanks to the algebraic
formulation. Thus a lot more can be said about the spectrum,
correlation functions, etc. of the quantum string theory based on these
models. Furthermore, it has been realized that the special geometries
described by these non-compact groups are relevant to gravitational
singularities such as black holes and cosmological Big Bang. For these reasons
this class of models has received considerable attention during the past
couple of years. It is hoped that through such solvable models new light will
be shed on unresolved gravitational issues, in string theory as well as
general relativity, such as  singularities, quantization and finiteness or
renormalizability in curved spacetime, the question of Euclidean-Minkowski
continuation, spectrum of low energy particles and excited string states in
the presence of curvature, etc..

\subsec{Heterotic Superstring in Curved Spacetime}

The other important question for string theory is the nature and content of
the low energy matter it is supposed to predict in the form of quarks,
leptons, gauge bosons, etc.. The new models have opened up the possibility of
heterotic superstring theories in four spacetime dimensions (with or without
additional compactified dimensions) \BShet .
This  is possible because the Virasoro central charge $c=26$ (or $c=15$ with
supersymmetry) condition can be satisfied in fewer dimensions provided the
space is curved. For example it has been possible to construct consistent
purely four dimensional heterotic string theories based on non-compact current
algebra cosets \IBhetero\IBerice .   We can now ask, what are the heterotic
models that can be constructed with the non-compact group method? Due to the
lack of space I will only give a couple of tables which are self-explanatory.
For details see the original literature. In these tables only the models in
purely four dimensions are given. The left movers are fully described in
Table-1 in the form of a supersymmetric Kazama-Suzuki coset, where $SO(3,1)_1$
represents four free Minkowski fermions. The right movers are described in
both tables. Table 1 contains the 4-dimensional coset for right movers
(connected with the left movers) while Table 2 gives the gauge group that is
needed in order to satisfy the $c_R=26$ condition. The contribution of the
gauge group to the central charge is given by $c_R(int)$ in Table 2.


$$\vbox   {    \tabskip=0pt \offinterlineskip \halign to 370pt  { \vrule# &
\strut # &
 #\hfil &\vrule# \ \ &  #\hfil &\vrule# \ \  &  #\hfil &\vrule# \tabskip =0pt
\cr \noalign {\hrule} && \# && left movers with N=1 SUSY && right movers &
\cr\noalign{\hrule} \noalign {\smallskip} &&$ 1 $&&${SO(3,2)_{-k}\times
SO(3,1)_{1}/ SO(3,1)_{-k+1}} $&
        &$  {SO(3,2)_{-k}/ SO(3,1)_{-k} }  $& \cr\noalign{\hrule} &&$ 2 $&&$
{SL(2,\IR)_{-k_1}\times SL(2,\IR)_{-k_2}\times SO(3,1)_1\over
        SL(2,\IR)_{-k_1-k_2+2}}\times \IR $&
        &$  {SL(2,\IR)_{-k_1}\times SL(2,\IR)_{-k_2}\over SL(2,\IR)_{-k_1-
k_2}}
 \times \IR   $& \cr\noalign{\hrule} &&$ 3 $&&$ \bl (SO(2,2)_{-k}\times
SO(3,1)_1/SO(2,1)_{-k+2}\br)\times \IR $&
       &$ \bl (SO(2,2)_{-k}/SO(2,1)_{-k}\br)\times \IR $& \cr\noalign{\hrule}
&&$ 4 $&&$ SL(2,\IR)_{-k}\times SO(3,1)_1\times \IR $&
        &$ SL(2,\IR)_{-k}\times \IR $& \cr\noalign{\hrule} &&$ 5 $&&$
{SL(2,\IR)_{-k_1}\times SL(2,\IR)_{-k_2}\times SO(3,1)_1 \over
          \IR^2 } $&
        &$ {SL(2,\IR)_{-k_1}\times SL(2,\IR)_{-k_2} / \IR^2 }   $&
\cr\noalign{\hrule} &&$ 6 $&&$ {SL(2,\IR)_{-k_1}\times SU(2)_{k_2}\times
SO(3,1)_1 / \IR^2} $&
        &$  {SL(2,\IR)_{-k_1}\times SU(2)_{k_2}/ \IR^2}  $&
\cr\noalign{\hrule} &&$ 7 $&&$ {(SL(2,\IR)_{-k}\times \IR^2\times SO(3,1)_1)/
\IR }  $&
        &$ {(SL(2,\IR)_{-k}\times \IR^2 )/ \IR }  $& \cr\noalign{\hrule} &&$ 8
$&&$ \IR^3\times \IR_Q\times SO(3,1)_1  $&&$ \IR^3\times \IR_Q   $&
\cr\noalign{\hrule} && \multispan5 Table 1. Current algebraic description of
left movers and right movers.\hfill & \cr\noalign{\hrule} }}$$
In item 8, $\IR_Q$ corresponds to a free boson with a background charge $Q$.


$$\vbox   {    \tabskip=0pt \offinterlineskip
\halign to 361pt  { \vrule# & \strut # &
 #\hfil &\vrule# \ \ &  #\hfil &\vrule# \ \  &  #\hfil &\vrule#
 \  & #\hfil &\vrule#
\tabskip =0pt
\cr \noalign {\hrule}
&& \# && conditions for $c_L=15 $ && $c_R(int)$ && gauge group, right movers &
\cr\noalign{\hrule}
\noalign {\smallskip}
&&$ 1 $&&$ k=5 $&&$ 11 $&&$ (E_7)_1\times SU(5)_1 $&
\cr\noalign{\hrule}
&&$ 2 $&&$ k_1-2={k_2-2\over 2}(-1+\sqrt {3k_2\over 3k_2-8})   $&
        &$ 13-\delta $&&$ \delta={12\over (k_1+k_2-4)(k_1+k_2-2)}  $&
\cr\noalign{\hrule}
&&$ 3 $&&$ k=3   $&&$ 11{1\over 2} $&
        &$ (E_7)_1\times SU(3)_1\times SU(2)_2\times U(1)_1  $&
\cr\noalign{\hrule}
&&$ 4 $&&$ k=8/3   $&&$ 13  $&&$ (E_8)_1\times SO(10)_1  $&
\cr\noalign{\hrule}
&&$ 5 $&&$ k_1={8k_2-20\over 3k_2-8},\ \ k_1,k_2>{8\over 3}  $&
        &$ 13 $&&$ (E_8)_1\times SO(10)_1 $&
\cr\noalign{\hrule}
&&$ 6 $&&$k_1={8k_2+20\over 3k_2+8}, \ k_2=1,2,3,\cdots $&
        &$ 13  $&&$ (E_8)_1\times SO(10)_1 $&
\cr\noalign{\hrule}
&&$ 7 $&&$ k={8/3} $&&$ 13  $&&$ (E_8)_1\times SO(10)_1  $&
\cr\noalign{\hrule}
&&$ 8 $&&$ Q^2={3\over 4} $&&$ 13  $&&$ (E_8)_1\times SO(10)_1  $&
\cr\noalign{\hrule}
&& \multispan7 Table 2. Conditions for $c_L=15$ and examples of symmetries
that give $c_R=26$. &
\cr\noalign{\hrule}
}}$$

It is encouraging to note that {\it the desirable low energy symmetries,
including $SU(3)\times SU(2)\times U(1)$, are contained in these curved
spacetime string models that have only four dimensions}. Also, the grand
unified gauge groups that emerge are the familiar and desirable ones. The
quark and lepton states, which come in color triplets and $SU(2)$ doublets,
are expected to emerge in several families. Compared to the popular approach
of four flat dimensions plus compactified dimensions, the gauge groups are
either the same or closely related. This gives the hope that the quark/lepton
spectrum of a curved purely four dimensional heterotic superstring that
describes the very early universe may be closely related to the quarks and
leptons that survive to the present times.

In trying to solve the puzzles of gravitational singularities and cosmology,
and those of the Standard Model with respect to the spectrum of matter (i.e.
quark/lepton families) and gauge bosons, we may hope that a complete string
theory in curved spacetime may guide us. For this reason I believe that it is
valuable to study in great detail the models presented in Table 1. These are
solvable models that should direct us toward a realistic unified theory.

Modular invariance for these models would be assured by the existence of a
Lagrangian formulation. Such a Lagrangian formulation was introduced in \BShet
. However, it has recently been pointed out in \grt\
that there is a global anomaly. To cancel the anomaly they proposed to
supersymmetrize both left and right movers. A simpler, and clearer version of
this proposal is the following: build supersymmetric gauged WZW Lagrangian
models for {\it both} right and left movers, as in \BShet , and then introduce
additional internal fermions to make up the deficit for $c_R=26$. This
modification would change the coset structure of the right movers: in Table 1
the left and right movers become the same, and in Table 2 the change of
$c_{int}$ modifies the internal groups slightly. Unfortunately, the authors of
\grt\ overlooked that their proposal creates another problem: it requires the
introduction of a time-like signature right-moving fermion (because of the
non-compact nature of the gauge subgroup) whose negative norm states cannot be
removed (since the right movers do not have overall superconformal symmetry).
Therefore, this proposal is defective and cannot be used. Instead, another
anomaly cancellation or modular in

 Note that if the time coordinate remains flat, and supersymmetric only for
left movers, the modified scheme would work without any problems. An important
example of such a model was already worked out in detail in \IBhet . To
specialize it to only 4 dimensions do as follows: on the left, take the
supersymmetrized time coordinate and one space coordinate to form flat
light-cone super coordinates with $c_L=3$, and take the supersymmetric,
left-right symmetric, $SL(2,\IR)_k/U(1)$ (at $k=8/3$ or $c=3k/(k-2)=12$) as two
more spacelike coordinates, to make altogether 4 coordinates. This gives
$c_L=15$. Furthermore, on the right, take 2 bosonic lightcone coordinates, plus
the right-moving supersymmetric $c=12$ mirror theory, and additional right
moving internal coordinates with $c_{int}=12$ to give $c_R=2+12+12=26$. The
4-dimensional geometry consists of two flat lightcone coordinates plus the dual
cigar/trumpet geometries of the $SL(2,\IR)/U(1)$ model (see below). The
computation of the spectrum of this
4-dimensional heterotic string model follows the path of ref.\IBhet\ with the
substitution of $(k=8/3, \ c=12)$ instead of $(k=3,\ c=9)$ used there. More
details will be given elsewhere.


\newsec{GEOMETRY OF THE MANIFOLD}

A gauged WZW model given by \action\actionone\ can be rewritten in the form of
a non-linear sigma model by choosing a unitary gauge that eliminates some of
the degrees of freedom from the group element, and then integrating out the
non-propagating gauge fields \BN\WIT . The remaining degrees of freedom are
identified with the string coordinates $X^\mu(\tau,\sigma)$. The resulting
action exhibits a gravitational metric $G_{\mu\nu}(X)$ and an antisymmetric
tensor $B_{\mu\nu}(X)$ at the classical level.

\eqn\eff{ S_{eff}=\int d^2\sigma (G^{\mu\nu}(X)\partial_\alpha
 X_\mu\partial^\alpha X_\nu
+ \cdots ) \ . }
At the one loop level there is also a dilaton $\Phi(X)$. These fields govern
the spacetime geometry of the manifold on which the string propagates.
Conformal invariance at the one loop level demands that they satisfy coupled
Einstein equations. Thanks to the exact conformal properties of the gauged
WZW model {\it the Einstein equations are automatically satisfied}. Therefore,
any of our non-compact gauged WZW models can be viewed as generating
automatically a solution of these equations. One only needs
to do some straightforward algebra to extract the explicit forms of
$G_{\mu\nu}, B_{\mu\nu}, \Phi$.

This algebra can be carried out by starting from the Lagrangian, such as in
\action , and has been done for all the models in four dimensions listed in
Table 1. The first case was $SL(2,\IR)/\IR$, which was interpreted by Witten
\WIT\ as the geometry of a 2D black hole. The higher dimensional cases yield
more intricate but singular geometries \BSthree\others\BShet\GIN\ .
In these manifolds there are geodesically complete patches that contain
singularities that hide behind a horizon, as well as patches that
contain bare singularities.
Although the Lagrangian method is straightforward, it has a number of
drawbacks. First, it yields the geometry only in a patch that is closely
connected to a particular choice of a unitary gauge. The remaining patches of
the global geometry can be recovered only in other unitary gauges and may have
no resemblance to the analytic form of the metric, dilaton, etc. in another
unitary gauge. To overcome this problem we have introduced global coordinates
\BSglo\ on the complete geometry. The global coordinates are gauge invariant.
The second problem with the Lagrangian method is that it yields the semi-
classical geometry up to one loop in an expansion in powers of $1/k$. However,
since the gauged WZW model is conformally exact one would rather obtain the
conformally exact geometry by using alternative methods.
 In the exact geometry some singularities are shielded by quantum
effects.
 It turns out that the
Hamiltonian method that utilizes the GKO construction solves both of these
problems simultaneously and yields an exact metric and dilaton to all orders
in $1/k$ \BSexa\SFET\BSslsu. More recently, the quantum effective action has
been constructed exactly, at least in the zero mode sector \BSeffaction\ for
any $G/H$ (see also \BST\TSEYT\ for the special case $SL(2)/\IR$) . The zero
mode sector is sufficient to extract the metric, dilaton and the antisymmetric
tensor that play a role in the low energy limit of string theory. From this
effective action it has finally been possible to extract the general exact
geometry for any coset $G/H$ \BSeffaction . However, although exact to all
orders in $1/k$, the known methods with the effective action still yield the
geometry in a patch instead of the global space. Therefore, in this paper we
concentrate on the algebraic Hamiltonian approach, which is really the main
basis for justifying the effective action.

With the Hamiltonian approach one can compute the
gravitational metric and dilaton backgrounds to all orders in the quantum
theory (all orders in the central extension $k$). We have managed to obtain
these quantities for bosonic, type-II supersymmetric, and heterotic string
theories. It turns out that the geometry of the heterotic and
type-II superstrings are obtained by deforming the geometry of the purely
bosonic string by definite shifts in the exact $k$-dependence. Therefore, it
is sufficient to first concentrate on the purely bosonic string. The following
relations have been proven for any $G/H$ model \BSexa\tseytlin : (i) For
type-II superstrings the conformally {\it
exact} metric and dilaton are identical to those of the non-supersymmetric
{\it semi-classical} bosonic model. (ii) The exact expressions for the
heterotic superstring are derived from their exact bosonic string counterparts
by shifting the central extension $k\to 2k-h$ (except for a different shift in
the overall $k$ factor). (iii) The combination $e^\Phi\sqrt{-G}$ is independent
of
$k$ and therefore can be computed in lowest order perturbation theory. Cases
2,5,6 in Table 1 are a bit more complicated because of the two central
extensions, but the results that relate the bosonic string to superstrings are
analogous. Case 6 is explicitly discussed in \BSslsu , and the others are
just analytic continuations of this one.

The main idea is the following. For the bosonic string the conformally exact
Hamitonian is the sum of left and right Virasoro generators $L_0+\bar
L_0$.
They may be written purely in terms of Casimir operators of $G$ and $H$ when
acting on a state $T(X)$ at the tachyon level. The exact dependence on the
central extension $k$ is included in this form by using the GKO formalism in
terms of currents. For example for the left-movers
 \foot{ $g$, $h$ are the Coxeter
numbers for the group and the subgroup. For the cases of interest in this
paper $g=d-1$, $h=d-2$ for $d\ge 3$, and $g=2,\ h=0$ for $d=2$. }

\eqn\lzer{\eqalign{&L_0 T=\bl({\D_G\over k-g}-{\D_H\over k-h}\br)T\cr
&\D_G \equiv Tr(J_G)^2, \qq \D_H \equiv Tr(J_H)^2\ ,\cr} }
The exact quantum eigenstate $T(X)=<X|Tachyon>$ can be analyzed in $X$-space.
Then the Casimir operators become Laplacians constructed as differential
operators in group parameter space (of dimension $dim( G)$). Consider a state
$T(X)$ which is
a singlet under the gauge group $H$ (acting simultaneously on left and right
movers)

\eqn\cond{(J_H +\bar J_H)\ T=0\ .}

\no
Because of the $dimH$ conditions $T(X)$ can depend only on
$d=dim(G/H)$ parameters, $X^\mu$ (string coordinates), which are
$H$-invariants constructed from group parameters (see below). The fact that
there are exactly $dim(G/H)$ such independent invariants is not immediately
obvious but it should become apparent to the reader by considering a few
specific examples. As discussed in \BSglo\ these are in fact the
coordinates that globally describe the sigma model geometry. Consequently,
using the chain rule, we reduce the derivatives in \lzer\ to only derivatives
with respect to the $d$ string coordinates $X^\mu$. In this way we can write
the
conformally exact Hamiltonian $L_0+\bar L_0$ as a Laplacian differential
operator in the global curved space-time manifold involving only the string
coordinates $X^\mu$. By comparing to the expected general form

\eqn\laplacian{ (L_0+\bar L_0)T={-1\over e^\Phi\sqrt{-
G}}\partial_\mu(e^\Phi\sqrt{-G}G^{\mu\nu}\partial_\nu T)}
for the singlet $T$, we read off the exact global metric and dilaton.

We have applied this program to all the models in Table 1 and obtained the
exact geometry to all orders in $1/k$. The large $k$ limit of our results
agree with the semi-classical computations of the Lagrangian method.
In the special case of two dimensions we also agree with another previous
derivation of the exact metric and dilaton for the $SL(2,\IR)/\IR$ bosonic
string \DVV .
 We summarize here the global and conformally exact results for the metric and
dilaton in the case of $SO(d-1,2)_{-k}/SO(d-1,1)_{-k}$ for d=2,3,4 \BSexa .
Due to the more complex expressions we refer the reader to the original
literature for the remaining cases \SFET\BSslsu . The group element $g$ for
$SO(d-1,2)/SO(d-1,1)$ can be parametrized as a $(d+1)\times (d+1)$ matrix in
the form

\eqn\group { g=\left ( \matrix {1 & 0 \cr 0 & ({1+a\over 1-
a})_{\alpha}{}^{\beta}\cr }
 \right  )  \ \left (\matrix {b  & (b+1) x^\beta \cr
-(b+1) x_\alpha  & (\eta_\alpha^{\ \beta} -(b+1) x_\alpha x^\beta) \cr } \right
),}

\no
where $b={1-x^2\over {1+x^2}}$. The $d$ parameters $x_\alpha$ and $d(d-1)/2$
parameters $a_{\alpha\beta}$ transform as a vector and an antisymmetric tensor,
respectively, under the Lorentz subgroup $H=SO(d-1,1)$ which acts on both sides
of the matrix as $g\rightarrow hgh^{-1}$. By considering the infinitesimal
left transformations $\d_L g=\e_L g$ we can read off the
generators that form an $SO(d-1,2)$ algebra for left transformations.

\eqn\leftgen{\eqalign{&J_{\alpha\beta}={1\over
2}(1+a)_{\alpha\alpha'}(1+a)_{\beta\beta'}
{\partial\over \partial a_{\alpha'\beta'}}\cr
 &J_{\alpha}=-{1\over 2}(1+x^2)\bl({1+a\over 1-a}\br)_{\alpha}{}^{\beta}
{\partial \over
\partial x^{\beta}} +{1\over 2}(1+a)_{\alpha\alpha'}(1+a)_{\beta'\g}x^{\g}
{\partial\over \partial a_{\alpha'\beta'}}\ .}  }

\no
If we consider instead the infinitesimal right transformations
$\d_R g=g \e_R$ we find the following expressions for the generators of right
transformations:

\eqn\rightgen{\eqalign{&\bar J_{\alpha\beta}=
-{1\over 2}(1-a)_{\alpha\alpha'}(1-a)_{\beta\beta'}
{\partial\over\partial a_{\alpha'\beta'}}-
x_{[\alpha}{\partial \over \partial x^{\beta]}}\cr
&\bar J_{\alpha}={1\over 2}(x^2-1){\partial\over \partial x^{\alpha}}-
x_{\alpha}x^{\beta} {\partial \over \partial x^{\beta}} -{1\over 2}(1-
a)_{\alpha\alpha'}(1-a)_{\g\beta'}x^{\g} {\partial\over \partial
a_{\alpha'\beta'}}\ .}}

\no
The $\bar J$ currents obey the same commutation rules as $J$ and moreover
commute with each other: $[J,\bar J]=0$.  The quadratic Casimirs for the
group and subgroup on either the left or the right are obtained by squaring
these currents. For the explicit expressions see \BSexa .

As argued above the global parametrization of the manifold is given in terms
of H-invariants, i.e. Lorentz invariants in the present case. In order to
obtain a diagonal metric on the manifold one must find $d$ convenient
combinations of these Lorentz invariants in $d$ dimensions. We give here the
basis that diagonalizes the semi-classical metric at large $k$. One of the
natural invariants already occurs in the construction of the group element for
every $d$, namely $b={1-x^2\over 1+x^2}$.

\subsec{Two dimensions}

For $d=2$ the antisymmetric tensor is Lorentz invariant
$a_{\alpha\beta}=a\epsilon_{\alpha\beta}$, and it is convenient to parametrize
$a=tanh(t)\ or\ coth(t)$. Then the global string coordinates can be taken as
$X^\mu=(t,b)$. Given all possible values for $(a,x^\alpha)$ the ranges of the
two
invariants cover the entire plane $-\infty<t,b<+\infty$.
The metric is given by the line element

\eqn\twometric{ds^2=2(k-2)\bl({db^2\over 4(b^2-1)}
-\b(b) {b-1\over b+1} dt^2\br),
\qq \b^{-1}(b)=1-{2\over k} {b-1\over b+1}\ . }

\no
For the dilaton the corresponding expression is

\eqn\twodilaton{\Phi=\ln\bl({b+1\over \sqrt{\b(b)}}\br)+const\ . }

\no
The scalar curvature for this metric is

\eqn\curv{R={2k\over k-2} {(k-2)b+k-4\over \bl((k-2)b+k+2\br)^2}\ .}

\no
The curvature is singular at $b=-(k+2)/(k-2)$, which is also where
$\beta(b)=\infty$. These are the properties of the exact 2d metric. The semi-
classical metric is obtained by taking the large $k$ limit, for which $\b=1$.
Then the singularity is at $b=-1$. Following Witten this singularity is
interpreted as a black hole while the horizon is at $b=1$. The signature of
the space is $(+-)$ or $(-+)$ depending on the region in the $(t,b)$ plane as
indicated in Fig. 2 of \BSglo . The signature is understood by examining
the semi-classical metric. To see the connection to the Kruskal coordinates
used by Witten let $b=1-2uv$ and $u^2=e^{2t}|b-1|/2$, $v^2=e^{-2t}|b-1|/2$.

If one considers the supersymmetric version of this string model, then
one can easily show that the metric is identical to the
semiclassical metric, and does not get renormalized for higher
orders of $1/k$ \BSexa . Therefore, in the supersymmetric case the
singularity is exactly at $b=-1$ since $\beta (b)=1$ and
$R=2/(k-2)(b+1)$.

There are other 2D cosets that one may consider, such as $SO(2,1)/SO(2)$
and $SO(3)/SO(2)$, that are analytic continuations of $SO(1,2)/SO(1,1)$.
The same procedure applied to them yields the same metrics in terms
of the global coordinates given above, but in different regions of the
manifold parametrized by $b$ and $u$, where $u=t^2$ is analytically
continued to negative values as well. The different regions have the
following
interpretations (see fig.2 in \BSglo ):
$SO(3)/SO(2)$= a cymbal for $-1<b\le 1$;
 $SO(2,1)/SO(2)$ = a cigar
when $b\ge 1$, and a trumpet when $b\le -1$; $SO(1,2)/SO(1,1)$ = region
outside of the horizon of a black hole when $b\ge 1$, region between
the horizon and singularity of the black hole when $-1<b<1$, bare
singularity region on the other side of the black hole when $b<-1$.

Properties of these geometrical spaces have been discussed in various
publications. In terms of the global coordinates defined here
a discussion of geodesics can be found in \BSglo . Furthermore, these
spaces have  ``duality" properties which correspond to a symmetry
that interchanges patches of the manifold without changing its
Hamiltonian. In the present case this corresponds to the operation
$b\rightarrow -b$. In terms of the original group parameters this is
generated by the inversion $x_\alpha \rightarrow x_\alpha /x^2 $. The
details are found in \BSglo\ and \BSexa . All of these properties
are shared by the higher dimensional manifolds of the following
sections,
but will not be discussed here due to lack of space. The interested
reader can consult the above references.

In the black hole case,
there are asymptotically flat  regions which are displayed by the change of
coordinates $b=\pm\cosh {2z_1\over\sqrt{2(k-2)}},\ t={z_0\over \sqrt{2k} }$.
For large $z_1\to\pm\infty$ and any $z_0$ the exact metric and dilaton have
the asymptotic forms

\eqn\asymtwo{ ds^2=dz_1^2-dz_0^2,\qquad \Phi=\sqrt{2\over k-2}|z_1| ,}
displaying a dilaton which is asymptotically linear in the space direction,
just like a Liouville field in 2d quantum gravity with a background charge.
Despite the flat metric there is no Poincar\'e invariance due to the linear
dilaton. Note that both the region outside the horizon ($b\to +\infty$) and
the naked singularity region ($b\to -\infty$) are asymptoticaly flat.

It was noted that the location of the singularity shifts as a function of $k$.
If one concentrates on the patches that represent the black hole, one concludes
that the quantum effects have shielded the singularity. This was first noted in
\BSexa\ and elaborated on more recently by other groups. In fact the same kind
of singularity shielding phenomenon is present in the quantum effects of all
other non-compact cosets as discussed and illustrated in \BSexa . Is this
significant for black hole physics? Are black holes eliminated by quantum
effects in string theory? For comparison, it is also beneficial to consider the
compact $SU(2)_k/U(1)$ at {\it positive} $k$ (as demanded by unitarity). The
geometry is given by the cymbal corresponding to the region $-1<b<1$ with a
{\it classical} singularity at the boundary $b=-1$. For the quantum metric,
curvature, etc. substitute  $k$ by $-k$ in the above expressions. The
singularity then moves inside the classical region and does not get shielded.
On the other hand, in

In my opinion one should concentrate on the wavefunction $T$ (rather than the
quantum corrected metric) and interpret it as a probability distribution. It is
then seen that the probability grows logarithmically in the vicinity of the
black hole, leading to the interpretation that the particle likes to spend a
lot of  time there. The logarithmic behavior is square integrable with an
appropriate measure given by $\int e^\Phi\sqrt{G}|T|^2$. Depending on physical
boundary conditions that are imposed, the particle may also tunnel to the
geometrical patch on the other side of the black hole, a phenomenon possible in
quantum  mechanics. Such phenomena still need to be interpreted.

\subsec {Three dimensions}

For $d=3$ the antisymmetric tensor is equivalent to a
pseudo-vector $a_{\alpha\beta}=\epsilon_{\alpha\beta\lambda}y^\lambda$, from
which we
construct two convenient invariants $v=2/(1+y^2)$ and $u=-v(x\cdot y)^2/x^2$,
which together with $b$ provide a basis for the string coordinates
$X^\mu=(v,u,b)$. Given all possible values taken by $(x^\alpha,y^\alpha)$ the
allowed
ranges for the invariants are

\eqn\rangethree{ \eqalign {(+-+)\ or\  (-++)\quad & \{b^2>1 \ \& \ uv>0\}, \cr
(++-)\quad &\{b^2<1\ \& \ uv<0\}, \quad except \quad 0<v<u+2<2 . } }
The 3d conformally exact metric is given by the line element \BSexa

\eqn\trimetric{ds^2=2(k-2)\bl(G_{bb} db^2 +G_{vv} dv^2 +G_{uu} du^2
+2G_{vu} dvdu\br)\ . }
where
\eqn\tridef{\eqalign{&G_{bb}={1\over 4(b^2-1)}\cr
&G_{vv}=-{\b(v,u,b)\over 4v(v-u-2)}
\bl({b+1\over b-1} + {1\over k-1} {u+2\over v-u-2}\br)\cr
&G_{uu}={\b(v,u,b)\over 4u(v-u-2)}
\bl({b-1\over b+1} - {1\over k-1} {v-2\over v-u-2}\br)\cr
&G_{vu}={1\over 4(k-1)} {\b(v,u,b)\over (v-u-2)^2}\ ,\cr}  }
and

\eqn\defb{\b^{-1}(v,u,b)=1+{1\over k-1} {1\over v-u-2}
\bl({b-1\over b+1} (u+2) -{b+1 \over b-1} (v-2) -{2\over k-1}\br)\ . }
The exact dilaton is

\eqn\tridilaton{{\Phi}=\ln \bl({ (b^2-1) (v-u-2)\over \sqrt{\b(v,u,b)}}\br)
+ \Phi_0\ , }

\no
In the large $k$ limit one obtains the global version of a semi-classical
metric derived in \BSglo\ using Lagrangian methods

\eqn\semicl{ {ds^2\over 2(k-2)}{\big\vert}_{k\to \infty}={db^2\over 4(b^2-1)}-
{1\over v-u-2}\bl({b+1\over b-1} {dv^2\over 4v}-{b-1\over b+1}{du^2\over 4u}\br
) }
The signature $(+-+),\ or \ (-++),\ or\ (++-)$ depends on the region and is
indicated in Fig-1 of \BSglo . A three-dimensional view of this metric is
given in Figs-4 of \BSglo . The surface is where the scalar curvature
blows up. This coincides with the location where the dilaton blows up in the
large $k$ limit as seen from the above expression. The space has two
topological sectors denoted by the sign of a conserved ``charge"
$\pm=sign(v(b+1))=sign(u(b-1))$. The sign never changes along geodesics. A
more intuitive view of the space is obtained in another set of coordinates for
the plus sector $(b,\lambda_+,\sigma_+)$ and the minus sector $(b,\lambda_-
,\sigma_-)$, which are given by $\lambda^2_\pm=\pm v(b+1)$ and
$\sigma^2_\pm=\pm u(b-1)$. Then the singularity surface is shown in Figs 3 of
\BSglo .
In the plus region the singularity surface has the topology of the double
trousers with pinches in the legs. In the minus region we have the topology of
two sheets that divide the space into three regions.

There are asymptotically flat regions that may be displayed by a change of
variables to $b=\pm \cosh {1\over \sqrt{3(k-2)}}(2z_1-z_0)$,
$u=(\pm)^\prime\cosh
 {1\over \sqrt{3(k-2)}}(-z_1+2z_0) \cosh^2z_2$, $v=(\pm)^\prime \cosh {1\over
\sqrt{3(k-2)}}(-z_1+2z_0) \sinh^2z_2$ (here $(\pm)^\prime$ is a set of signs
indepedent than $\pm$ ). For large values of $z_1\to\pm\infty$,
and finite values of $(z_0,z_2)$, the semiclassical metric and dilaton take
the form

\eqn\asymthree{ ds^2=-dz_0^2+dz_1^2+dz_2^2, \qquad
\Phi=\sqrt{6\over k-2}|{5\over 3}z_1-{4\over 3}z_0|,}
showing that the dilaton is linear in a space-like direction $z_1'={5\over
3}z_1-{4\over 3}z_0$ in the asymptotically flat region. Then $z_1'$ behaves
just like a Liouville field, while the Lorentz transformed $z_0'={5\over
3}z_0-{4\over 3}z_1$ is a time coordinate, and the diagonal metric is
rewritten as $ds^2=-(dz_0')^2+(dz_1')^2+dz_2^2$. The exact metric is not flat
when only $|z_1|$ is large. To display its asymptotically flat region one
requires somewhat different coordinates.

\subsec{Four dimensions}

For $d=4$ one can construct the Lorentz invariants

\eqn\testates{x^2\ ,\qq z_1={1\over 4}Tr(a^2)\ ,\qq
z_2={1\over 4}Tr(a^* a)\ , \qq z_3=xa^2x/x^2 \ ,}

\no
where $a^*_{\alpha\beta}={1\over
2}\e_{\alpha\beta\alpha'\beta'}a^{\alpha'\beta'}$ is the dual of
$a_{\alpha\beta}$. However, the semi-classical metric  is diagonal for a
different set of four invariants $X^\mu=(v,u,w,b)$ given by

\eqn\relat{\eqalign{&b={1-x^2 \over 1+x^2}\ , \qq
u={1+z_2^2+2(z_1 -z_3) \over 1-2 z_1-z_2^2}\cr
&v={1+z_1 +\sqrt{z_1^2 +z_2^2} \over 1-z_1 -\sqrt{z_1^2 +z_2^2}}\ ,\qq
w={1+z_1 -\sqrt{z_1^2 +z_2^2} \over 1-z_1 +\sqrt{z_1^2 +z_2^2}}\ .} }

\no
To find the ranges in which the above global coordinates take their values we
consider a Lorentz frame that can cover all possibilities without loss of
generality. First we notice that by Lorentz transformations the antisymmetric
matrix $a_{\alpha\beta}$ can always be transformed to a block diagonal
matrix with the non-zero elements

\eqn\blo{a_{01}=\tanh t\ {\rm or}\ \coth t\ ,\qq a_{23}=\tan \phi\ .}

\no
Then, using \relat\ one can deduce the form of the global variables:
$v=\pm \cosh 2t$, $w=\cos 2\phi$, and $u={1\over x^2}
\bl(w(x_0^2-x_1^2)-v(x_2^2 +x_3^2)\br)$. Therefore the string variables
can take values in the following regions with the signature in the
$(v,u,w,b)$ basis

\eqn\range{\eqalign{
& (-+++):\quad b^2>1,\quad \{-1<w<u<1<v\ \ or \ \ v<-1<u<w<1 \cr
& \hskip 5cm or\ -1<w<1<u<v \},\cr
& (+-++):\quad b^2>1, \quad \{-1<w<1<v<u\ \ or \ \ u<v<-1<w<1 \}\cr
& (+++-):\quad b^2<1, \quad \{u<w<11<v \ or \ v<-1<w<u \ or \ v<u<-1<w<1\}. }}

With this set of coordinates we compute the conformally exact dilaton and
metric as before. The dilaton field is

\eqn\fourdilaton{\Phi=\ln \bl({(b^2-1)(b-1)(v-u)(w-u)\over
\sqrt{\b(b,u,v,w)}}\br) +\Phi_0\ . }
and the metric is given by

\eqn\fourmetric{\eqalign{ds^2=2(k-3)
&\bl(G_{bb} db^2 +G_{uu} du^2 +G_{vv} dv^2 +G_{ww} dw^2\cr
&+2 G_{uv} du dv +2 G_{uw} du dw +2 G_{vw} dv dw\bl)\ ,\cr} }
where

\eqn\fourmet{\eqalign{&G_{bb}={1\over 4(b^2-1)}\cr
&G_{uu}={\b(b,u,v,w)\over 4(u-w)(v-u)}\biggl({b-1\over b+1}-{1\over k-2}
{(v-w)^2\over {(v-u)(u-w)}} (1-{1\over k-2} {b+1\over b-1})\biggr)\cr
&G_{vv}=-{(v-w)\b(b,u,v,w)\over 4(v^2-1)(v-u)}\biggl({b+1\over b-1}-{1\over k-
2} {1\over (v-u)(u-w)}\bl[1-u^2+\cr
 &\quad +({b+1\over b-1})^2 (v-u)(v-w) +{1\over k-2}
{b+1\over b-1} {(1+v^2)(u+w)-2v(1+uw)\over v-w}\br]\biggr)\cr
&G_{ww}={(v-w)\b(b,u,v,w)\over 4(1-w^2)(u-w)}\biggl({b+1\over b-1}-{1\over k-2}
{1\over (v-u)(u-w)}\bl[1-u^2+\cr
 &\quad +({b+1\over b-1})^2 (u-w)(v-w)-{1\over k-2}
{b+1\over b-1} {(1+w^2)(u+v)-2w(1+uv)\over v-w}\br]\biggr)\cr
&G_{uv}={\b(b,u,v,w)\over 4(k-2)(v-u)^2}\bl(1-{1\over k-2} {b+1\over b-1}
{v-w\over u-w}\br)\cr
&G_{uw}={\b(b,u,v,w)\over 4(k-2)(u-w)^2}\bl(1-{1\over k-2} {b+1\over b-1}
{v-w\over v-u}\br)\cr
&G_{vw}={1\over (k-2)^2} {b+1\over b-1} {\b(b,u,v,w)\over
4(v-u)(u-w)}\ ,\cr} }

\no
and the function $\b(b,u,v,w)$ is defined by

\eqn\defbb{\eqalign{ & \b^{-1}(b,u,v,w)=1+{1\over k-2} {(v-w)^2
\over (v-u)(w-u)} \biggl({b+1\over b-1}+{b-1\over b+1} {1-u^2\over (v-w)^2}\cr
&\quad +{1\over k-2}\bl({vw+u(v+w)-3 \over (v-w)^2} -({b+1\over b-
1})^2\br)\biggr) +{2\over (k-2)^3} {b+1\over b-1} {vw-1 \over (v-u)(u-w)}\
.\cr} }

The large $k$ limit of these expressions reduce to the semiclassical dilaton
and metric that follow from the Lagrangian approach

\eqn\semetr{\eqalign{{ds^2\over 2(k-2)} {\big \vert}_{k\to \infty}=
&{db^2\over 4(b^2-1)}+{b-1\over b+1} {du^2\over 4(v-u)(u-w)}\cr
&+{b+1\over b-1}(v-w)\bl({dw^2\over 4(1-w^2)(u-w)}
-{dv^2\over 4(v^2-1)(v-u)}\br)\ .\cr} }

We can see that the signature of the semiclassical metric for different ranges
of the parameters \range\ is precisely as required by the group parameter
space which led to \range . However, for the exact metric $\beta(u,v,w,b)$ must
remain positive to keep $-det(G)$ positive. This implies that part of the
regions in \range\ are screened out by quantum effects for the exact geometry.
This screening phenomenon is true for every dimension $d=2,3,4$ and the
screened regions must be interpreted in the quantum theory as tunneling or
decay regions for probability amplitudes (such as the tachyon wavefunction).
Under any circumstances the manifold cannot go outside of the range \range\
dictated by the group theory.

As in the previous $d=2,3$ cases, we can check that our explicit expressions
for the dilaton and metric give the $k$-independent combination $\sqrt{-
G}e^{\Phi}$. Therefore this quantity takes the same value for either the
exact metric and dilaton or the semiclassical metric and dilaton. Since it
is unrenormalized by quantum effects (other than one loop), it may be
computed in lowest order perturbation theory. This combination appears in the
d'Alembertian and is also closely related to the integration measure in the
path integral. Through group theoretical arguments given in
\BSthree\BShet\ it was possible to guess that this combination should
remain unrenormalized by quantum effects. Similar to the $d=2,3$ cases the 4d
manifold has an asymptotically flat region, but it will not be discussed here.

\subsec{Particle and String Geodesics}

Having global coordinates and a global geometry is not sufficient to get a
feeling of the geometry, one also needs to know the behavior of the
geodesics. However, for the complicated metrics that are displayed above the
geodesic equation seems to be completely unmanageable. Fortunately, we have
developed a procedure that relies on group theory and managed to solve for all
particle geodesics. The trick is to take advantage of the fact that the global
coordinates are gauge invariant under $H$-transformations. Then we may solve
the equations of motion for the group element $g$ in any gauge, and use
{\it the solution} to construct the $H$-invariant combinations that form the
global coordinates of the geometry. In fact, there is an axial gauge in which
$g$ is solved easily \BSglo . For a point particle (string shrunk to a
point) it is given as a function of proper time

\eqn\groupel{ g(\tau)=e^{\alpha\tau} g_0 e^{(p-\alpha)\tau}, }
where $g_0$ is a constant group element at initial proper time $\tau$, and
$\alpha , p$ are constant matrices in the Lie algebras of $H$ and $G/H$
respectively. The equations of motion require that these constants satisfy a
constraint

\eqn\constraint{ (g_0(p-\alpha)g_0^{-1})_H+\alpha=0\ ,}
where the subscript $H$ implies a projection to the Lie algebra of $H$. This
solution applies to any group and subgroup. As shown in \BSglo\ the
standard geodesics equations for the geometries displayed above are
automatically solved when the $H$-invariants are constructed from the solution
\groupel\constraint . In this way all light-like, space-like and time-like
geodesic solutions are obtained.

With the point geodesics at hand we have discovered a number of additional
interesting properties about the $d=2,3,4$ manifolds \BSglo\ which
generalize to other non-compact gauged WZW models as well. The most striking
feature is that the manifolds that are pictured in the figures have many
copies and the complete manifold must include all the copies. The gauge
invariant coordinates (e.g. $(b,t)$ for $d=2$) are not sufficient to fully
describe the structure. There are additional {\it discrete} gauge invariants
constructed from the group element $g$ that label the copies of the manifold.
This can be seen easily in our examples since the gauge subgroup is just the
Lorentz group and its properties are well known. In this case the invariants
are Lorentz dot products constructed from a vector $x^\alpha$ and a tensor
$a^{\alpha\beta}$. Let us consider the invariant $b=(1-x^2)/(1+x^2)$, say in
the
region $x^2>0$. It is known that the time component $x^0$ could be either
positive or negative and that a Lorentz transformation cannot change this
sign. Therefore, the sign of $x^0$ is a discrete gauge invariant which does
not show up in the metric or dilaton that characterized the manifolds
discussed above. However, the model as a whole knows about this sign through
the
group element $g$. Such discrete invariants are present in every
{\it non-compact} gauged WZW model and they label copies of the manifolds
described above. We may then ask whether these copies communicate with each
other? The answer is yes, they do, and this can be seen by following the
behaviour of a particle geodesic. The full information about the particle
geodesic is contained in the solution for $g$ in \groupel\constraint . From
this it can be verified that at the proper time that a particle touches a
curvature singularity the discrete invariant switches sign and then the
particle continues its journey smoothly from one copy of the manifold to the
next. For example, in the 2d black hole case this happens for a time-like
geodesic (i.e. massive particle) in a finite amount of proper time (on the
other hand, a light-like geodesic takes an infinite amount of proper time to
reach the singularity and therefore ends its journey without changing copies
of the manifold). This behavior is present in all non-compact models in this
paper as well as in other models (e.g. we have verified it in the $
SL(2,\IR)\times SU(2)/\IR^2 $ model). It is reminiscent of the
Reissner-Nordtsr\"om black hole in which geodesics move on to other worlds. The
difference is that in our case this happens at the singularity itself. When
quantum corrections are included and the exact metric considered, then the
singularity and the transition to other worlds no longer seem to be at the
same place; at least this is the case for the 2d black hole.
The spectrum of the discrete invariant depends on the group
representation and therefore one expects different numbers of copies in
different quantum states. The number of copies is infinite for quantum states
with non-fractional quantum numbers, which is typical in unitary
non-holomorphic representations of non-compact groups. When the number of
copies is infinite the particle can never come back to the same world, but
for a finite number of copies the particle returns to the original world by
emerging from a white singularity.

So far we have discussed particle geodesics that correspond to a string
collapsed to a single point. We may also investigate string geodesics in the
same manifolds. That is, we are also interested in solutions for the strings
moving in curved spacetime, just like one has a complete solution in flat
spacetime in terms of harmonic oscillator normal modes. This problem has been
solved in principle for the non-compact gauged WZW models in \BSthree .
There the solution for the group element $g(\tau,\sigma)$ has been obtained
explicitly in terms of normal modes. This is the analog of \groupel\ above.
There remains to construct the appropriate dot products to form the
invariants, which in turn are the solutions to the string geodesics. This last
part has not yet been performed explicitly, but it is only a matter of
straightforward algebra of the kind performed for the particle geodesics in
\BSglo . This procedure gives all the solutions in curved spacetime and
can answer questions of the type ``what happens when a string falls into a
black hole ?"

\subsec{Duality}

Due to  lack of space we have not covered other interesting topics such as
duality properties of these manifolds. It was shown in \BSthree\BSglo\
that there is a dynamical duality that generalizes the $R\rightarrow 1/R$
duality properties of conformal field theories based on tori. This is related
to the axial/vector duality that is present in the 2d black hole. It was shown
in \BSglo\ that the duality transformation is equivalent to an inversion
in group parameter space $(x_\alpha , a_{\alpha\beta})$ given in \group . This
inversion generates discrete jumps for the group parameter that corresponds to
interchanging different patches of the geometrical manifold. For details the
reader is referred to \BSglo . This duality property is closely related to
mirror symmetry of the kind discussed for Calabi-Yau manifolds, as will be
explained elsewhere. The duality symmetry mentioned here is different than the
one discussed in recent months by Verlinde, Giveon, Rocek, and others.

\newsec{ CONCLUSIONS }

We have only scratched the surface of the subject of non-compact gauged WZW
models. We have shown that this approach is very useful for learning about
strings in curved spacetime that may be relevant for the early part of the
Universe. It is during this era that string theory should be relevant and it is
during this era that the matter we know was formed. Therefore, in trying to
solve the puzzles of the Standard Model with respect to the spectrum of matter
and gauge bosons we may hope that a string theory in curved spacetime may
guide us. For this reason I believe that it is valuable to study in great
detail the models presented in Table 1. These are solvable models that should
direct us toward a realistic unified theory.

I did not have the space to discuss a number of interesting results.
Among these I want to mention the construction of the effective
quantum action to all orders in $k$, at least in the zero mode sector
\BSeffaction . This method generates the same exact conformal
metric and dilaton as the Hamiltonian approach, but in addition it also
gives the exact antisymmetric tensor $B_{\mu\nu}(X)$ (the axion). The
general results for all gauged WZW models are provided in \BSeffaction\
\foot { See also \TSEYT\ for the SL(2,R) case, with a partial discussion
of possible non-local terms for the higher string modes. However,
his treatment is not gauge invariant as noted in \BSeffaction .}.

Other important results are the computation of conformally
exact quantities for supersymmetric and heterotic strings.
These are obtained by a simple substitution of $k$ by a shifted
value of $k$ in the semiclassical or exact results of the purely
bosonic quantities. The prescription is derived in \BSexa\
and applied to a number of cases there and elsewhere \BSslsu .

There are many mathematical problems of interest. The geometries,
the duality properties, the unitary representations of non-compact groups
restricted to the appropriate subgroup, etc. are all problems that either
have not been studied, or require a lot more research. From
the point of view of physics, we are only at the beginning
of our understanding of strings in curved spacetimes and many
interesting results can be expected in the future.


\listrefs
\end